\documentclass[runningheads]{llncs}

\usepackage{cite}
\usepackage{amssymb}
\usepackage{amsmath}

\usepackage{graphicx}
\usepackage[caption=false]{subfig}

\usepackage{color}
\usepackage[dvipsnames]{xcolor}
\usepackage[colorlinks=false, linkcolor=blue, urlcolor=blue]{hyperref}
\newcommand{\identity}{\mathbf{I}}
\newcommand{\laplacian}{\mathbf{L}}
\newcommand{\adjacency}{\mathbf{A}}
\newcommand{\degree}{\mathbf{D}}
\newcommand{\eigvecs}{\mathbf{U}}
\newcommand{\eigvals}{\mathbf{\Lambda}}
\newcommand{\vertices}{\mathcal{V}}
\newcommand{\edges}{\mathcal{E}}

\definecolor{mustard}{RGB}{223, 175, 54}
\definecolor{baby blue}{RGB}{170, 223, 255}

\begin{document}
\title{Interpretation of Brain Morphology in Association to Alzheimer's Disease Dementia Classification Using Graph Convolutional Networks on Triangulated Meshes}
\titlerunning{Interpretation of Brain Morphology in Association to Alzheimer's Disease}
%
\author{Emanuel Azcona\inst{1, 4}\orcidID{0000-0002-1896-4823} \and Pierre Besson \inst{2, 4}\orcidID{0000-0002-0591-7542} \and Yunan Wu \inst{1, 4}\orcidID{0000-0001-6980-9746} \and Arjun Punjabi \inst{1, 4}\orcidID{0000-0003-2770-3112} \and Adam Martersteck \inst{3, 4}\orcidID{0000-0001-6997-9390} \and Amil Dravid \inst{1, 4}\orcidID{0000-0001-6007-0690} \and Todd B. Parrish \inst{3, 4}\orcidID{0000-0002-1184-1572} \and S. Kathleen Bandt \inst{2, 4}\orcidID{0000-0002-1525-2775} \and Aggelos K. Katsaggelos \inst{1, 4}\orcidID{0000-0003-4554-0070} for the Alzheimer’s Disease Neuroimaging Initiative\thanks{Data used in preparation of this article were obtained from the Alzheimer’s Disease Neuroimaging Initiative (ADNI) database (\url{adni.loni.usc.edu}). As such, the investigators within the ADNI contributed to the design and implementation of ADNI and/or provided data but did not participate in analysis or writing of this report. A complete listing of ADNI investigators can be found at: \url{http://adni.loni.usc.edu/wp-content/uploads/how_to_apply/ADNI_Acknowledgement_List.pdf}}}
\authorrunning{E. Azcona \emph{et al.}}
%
\institute{Image and Video Processing Laboratory, Department of Electrical and Computer Engineering, Northwestern University, IL, USA, \url{https://ivpl.northwestern.edu}
\and
Advanced NeuroImaging and Surgical Epilepsy (ANISE) Lab,\\Northwestern Memorial Hospital, IL, USA, \url{https://anise-lab.com/}
\and
Neuroimaging Laboratory, Department of Radiology, Northwestern University, IL, USA, \url{neuroimaging.northwestern.edu}
\and
Augmented Intelligence in Medical Imaging, Northwestern University, IL, USA}
\maketitle              
%
\begin{abstract}
    We propose a mesh-based technique to aid in the classification of Alzheimer's disease dementia (ADD) using mesh representations of the cortex and subcortical structures. Deep learning methods for classification tasks that utilize structural neuroimaging often require extensive learning parameters to optimize. Frequently, these approaches for automated medical diagnosis also lack visual interpretability for areas in the brain involved in making a diagnosis. This work: (a) analyzes brain shape using surface information of the cortex and subcortical structures, (b) proposes a residual learning framework for state-of-the-art graph convolutional networks which offer a significant reduction in learnable parameters, and (c) offers visual interpretability of the network via class-specific gradient information that localizes important regions of interest in our inputs. With our proposed method leveraging the use of cortical and subcortical surface information, we outperform other machine learning methods with a 96.35\% testing accuracy for the ADD vs. healthy control problem. We confirm the validity of our model by observing its performance in a 25-trial Monte Carlo cross-validation. The generated visualization maps in our study show correspondences with current knowledge regarding the structural localization of pathological changes in the brain associated to dementia of the Alzheimer's type.
    \keywords{Graph convolutional networks  \and Alzheimer's disease classification \and triangulated meshes \and neural network interpretability.}
\end{abstract}
\section{Introduction}
Alzheimer's disease dementia (ADD) is a clinical syndrome characterized by progressive amnestic multidomain cognitive impairment \cite{McKhann2011}. The causative underlying pathology is Alzheimer’s disease (AD), defined as the co-occurrence of neurofibrillary tangles and amyloid-beta plaques. Globally, the number of individuals living with AD is expected to reach 1 out of 85 people by the year 2050 \cite{Brookmeyer2007}. Automated methods for the computer-aided clinical diagnosis of ADD has been an area of interest in the medical imaging community for  the development of assistive tools aiding in the visual inspection of structural information captured by magnetic resonance imaging (MRI).

Previous studies in the neuroanatomical pathologies of AD have demonstrated correlations in cortical folding pattern \cite{ono1990atlas} and different neurodegenerative pathologies. Specific patterns of atrophy in the cortex and subcortical structures have been linked to AD \cite{Kalin2017a, Liu2012a}. For example, \cite{ono1990atlas} discusses a potential to focus on high variability in association cortices like the intermediate sulcus of Jensen. As \cite{Pacheco2015a} also points out, widespread cortical thinning and a greater rate of atrophy is present in temporal lobe regions, primarily the left parahippocampal gyrus, for subjects with AD. Furthermore, Jong \emph{et al.} \cite{DeJong2008StronglyStudy} discuss irregularities like reduced putamen and thalamus volumes for subjects with AD. In studies such as ADNI, it is common to find bias towards more left-sided atrophy because of the verbal language tests given to assess memory function \cite{Derflinger2011}. For example, if asymmetrical atrophy of the language network is more prominent, subjects may perform worse on verbal tests and be diagnosed with dementia earlier.

Machine learning (ML) methods have been a growing area of interest in the automated clinical diagnosis for ADD. \cite{ZHANG2012895, LIU2014466, beheshti2017classification} discuss the use of support vector machines (SVMs) in unimodal and multimodal imaging pipelines for the automated classification of ADD using MRI, PET, and cerebrospinal fluid (CSF). In \cite{Li2014a, arjun2019plosone}, the use of MRI and PET imaging in multimodal convolutional neural networks (CNNs) for ADD diagnosis is discussed. SVM-based approaches, like those used in \cite{ZHANG2012895, LIU2014466, beheshti2017classification}, have historically been hard to interpret, expensive to train, and often serve as the logical choice only when there is enough domain expertise to construct meaningful kernels. Multimodal volumetric CNNs like \cite{arjun2019plosone}, often require a lot of memory and frequently are limited to smaller-batch operations or using lower resolution 3D volumes.

Motivated by 3D object detection via surfaces \cite{Masci2015}, cortical and subcortical irregularities correlated with ADD, our work uses mesh manifolds of the cortex and subcortical structures in the diagnosis of ADD. Our technique leverages a reduction in computational complexity offered by \cite{Defferrard2016}. In \cite{PARISOT2018117}, Parisot \emph{et al.} leverage this work from \cite{Defferrard2016} to make similar predictions for Alzheimer's disease and Autism using graph convolutional networks (GCNs) on ADNI/ABIDE subject population graphs. In \cite{ranjan2018generating}, their convolutional mesh autoencoder (CoMA) framework uses the same GCN basis from \cite{Defferrard2016} on human face surface meshes to generate new meshes from a learned distribution conditioned on facial expression labels. Their network is also able to reconstruct input meshes from compressed 8-dimensional representations with a 50\% reduction in reconstruction error, while using 75\% fewer parameters than volumetric models that operate on voxels.

The interpretability of results from ML models has remained an open issue in highlighting regions of interest (ROI) in relation to classification decisions. In this paper we demonstrate that it is possible to (1) extract meaningful surface meshes of the cortex and subcortical structures, (2) achieve accurate predictions for the clinical binary classification of ADD using meshes, (3) extract class-discriminative localization maps for interpretable ROI, and (4) reduce the number of learnable parameters.
\section{Methods}
Data used in the preparation of this article were obtained from the Alzheimer’s Disease Neuroimaging Initiative (ADNI) database (\url{https://adni.loni.usc.edu}). The ADNI was launched in 2003 as a public-private partnership, led by Principal Investigator Michael W. Weiner, MD. The primary goal of ADNI has been to test whether serial magnetic resonance imaging (MRI), positron emission tomography (PET), other biological markers, and clinical and neuropsychological assessment can be combined to measure the progression of mild cognitive impairment (MCI) and early Alzheimer’s disease (AD).

\subsection{Localized Spectral Filtering on Graphs}
Spectral-based graph convolution methods inherit ideas from a graph signal processing (GSP) perspective as described by \cite{wu2019comprehensive}. Like \cite{Defferrard2016}, our work focuses on using undirected graphs defined by a finite set of vertices, $\vertices$, with $N = |\vertices|$ vertices, and a corresponding set of edges, $\edges$, with scalar edge weights, $e_{ij} = e_{ji} \in \edges$, which are stored in the $i^{th}$ rows and $j^{th}$ columns of the adjacency matrix, $\adjacency \in \mathbb{R}^{N \times N}$. A graph's node attributes are defined using the node feature matrix $\mathbf{X} \in \mathbb{R}^{N \times F}$ where each column, $\mathbf{x}_i \in \mathbb{R}^N$, represents the feature vector for a particular shared feature across each of the vertices, $v_i \in \vertices$.

A great emphasis in GSP is placed on the normalized graph Laplacian, $\laplacian = \identity_{N} - \degree^{-1/2}\adjacency\degree^{-1/2}$, where $\identity_N$ is the identity matrix and $\degree_{ii} = \sum_{j}\adjacency_{ij}$ is the diagonal matrix of node degrees. $\laplacian$ can be factored via the eigendecomposition: $\laplacian = \eigvecs\eigvals\eigvecs^T$, where $\eigvecs \in \mathbb{R}^{N \times N}$ is the complete set of orthonormal eigenvectors for $\laplacian$ and $\eigvals = diag\left(\left[\lambda_0, \dots, \lambda_{N-1}\right]\right) \in \mathbb{R}^{N \times N}$ is the corresponding set of eigenvalues. Given a spectral filter, $g_\theta$, defined in the graph's Fourier space \cite{shuman2013emerging} as a polynomial of the Laplacian, $\laplacian$, and $\eigvecs$'s orthonormality, we can filter $\mathbf{x}$ via multiplication s.t.
    \begin{equation}
        g_\theta *_\mathcal{G} \mathbf{x} = g_\theta(\laplacian)\mathbf{x} = g_\theta\left(\eigvecs\eigvals\eigvecs^T\right)\mathbf{x} = \eigvecs g_\theta(\eigvals) \eigvecs^T \mathbf{x} \label{eq:spectral_conv},
    \end{equation}
where $\theta \in \mathbb{R}^N$ are the parameters of the filter $g_\theta$ and $*_\mathcal{G}$ is the spectral convolution operator notation borrowed from \cite{Defferrard2016}. Furthermore, $\eigvecs^T\mathbf{x}$ is the \textit{graph Fourier transform} (GFT) of the graph signal $\mathbf{x}$, $g_\theta(\eigvals)$ is a filter defined using the spectrum (eigenvalues) of the normalized Laplacian, $\laplacian$, and the left-sided multiplication with $\eigvecs$ is the \textit{inverse}-GFT (IGFT). In this context, convolution is implicitly performed by using the duality property of the Fourier transform s.t. a spectral filter is first multiplied with the GFT of a signal, and then the IGFT of their product is determined.
 
 Our approach uses Chebyshev polynomials of the first kind \cite{Defferrard2016, garfken67:math} to approximate $g_\theta$ using the graph's spectrum s.t.
    \begin{equation}
        g_\theta(\tilde{\laplacian}) = \sum_{k=0}^{K-1} \theta_k T_k(\tilde{\laplacian}) \label{eq:cheb_poly},
    \end{equation}
for the scaled Laplacian $\tilde{\laplacian} = \frac{2 \laplacian}{\lambda_{max}} - \identity_N$, where $\lambda_{max}$ is the largest eigenvalue in $\eigvals$, and $K$ can be interpreted as the kernel size. Chebyshev polynomials of the first kind are defined by the recurrence relation, $T_k(\tilde{\laplacian}) = 2\tilde{\laplacian} T_{k-1}(\tilde{\laplacian}) - T_{k-2}(\tilde{\laplacian})$ where $T_0(\tilde{\laplacian}) = \identity$ and $T_1(\tilde{\laplacian}) = \tilde{\laplacian}$ as shown in \cite{Defferrard2016}.
\subsection{Mesh Extractions of Cortical \& Subcortical Structures} \label{sec:graph_extract}
Using FreeSurfer v6.0 \cite{Fischl2012}, all MRIs were denoised followed by field inhomogeneity correction, and intensity and spatial normalization. Inner cortical surfaces (interface between gray and white matter) and outer cortical surfaces (CSF/gray matter interface) were extracted and automatically corrected for topological defects. Additionally, seven subcortical structures per hemisphere were segmented (amygdala, nucleus accumbens, caudate, hippocampus, pallidum, putamen, thalamus) and modeled into surface meshes using SPHARM-PDM (\url{https://www.nitrc.org/projects/spharm-pdm}).

Surfaces were inflated, parameterized to a sphere, and registered to a corresponding spherical surface template using a rigid-body registration to preserve the cortical \cite{Fischl2012} and subcortical \cite{BESSON2014283} anatomy. Surface templates were converted to meshes using their triangulation schemes. A scalar edge weight, $e_{ij}$, was assigned to connect vertices $v_i$ and $v_j$ using their geodesic distance, $\psi_{ij}$, s.t.
    \begin{equation}
        e_{ij}= e_{ji} = \frac{1}{\sigma\sqrt{2\pi}}e^{-\frac{1}{2}\left(\frac{\psi_{ij}}{\sigma}\right)^2} \label{eq:edge_weights}.
    \end{equation}

Surface templates were parcellated using a hierarchical bipartite partitioning of their corresponding mesh. Starting with their initial mesh representation of densely triangulated surfaces, spectral clustering was used to define two partitions. These two groups were then each separated yielding four child partitions, and this process was repeated until the average distance across neighbor partitions was below $2.5$ mm. For each partition, the central node was defined as the node whose centrality was highest and the distance across two partitions was defined as the geodesic distance (in mm) across the central vertices. Two partitions were neighbors if at least one node in each partition were connected. Finally, partitions were numbered so that partitions $2i$ and $2i+1$ at level $L$ had the same parent partition $i$ at level $L-1$. Therefore, for each level a graph was obtained s.t. the vertices of the graph were the central vertices of the partitions and the edges across neighboring vertices were weighted as in Eq. \ref{eq:edge_weights}. This serves as an improvement upon \cite{Defferrard2016} to ensure that no singleton is ever produced by pooling operations for the cortex and subcortical structures. At the finest level, meshes had a total of $47,616$ vertices: $32,768$ vertices for the cortex and $14,848$ vertices to represent the subcortical structures.

Vertex features were defined as the Cartesian coordinates of the surface vertices in the subjects’ native space registered to the surface templates. This can create issues if the original scans are not registered to the same template, as was also done by Ranjan \emph{et al.} in \cite{ranjan2018generating}. Similar studies, like that of Gutiérrez-Becker and Wachinger \cite{gutierrez2019learning}, implement ``rotation network'' modules as the first few layers of their neural network (NN) architecture to aid in correcting and aligning their samples to a common template. Performing our template registration as an additional preprocessing step reduces the complexity of our NN architecture and eliminates the need of incorporating an ``alignment'' term to our cost function to optimize later, as was needed in \cite{gutierrez2019learning}.

\begin{figure}[!htb]
        \centering
        \subfloat[]{
            \includegraphics[width=.24\linewidth]{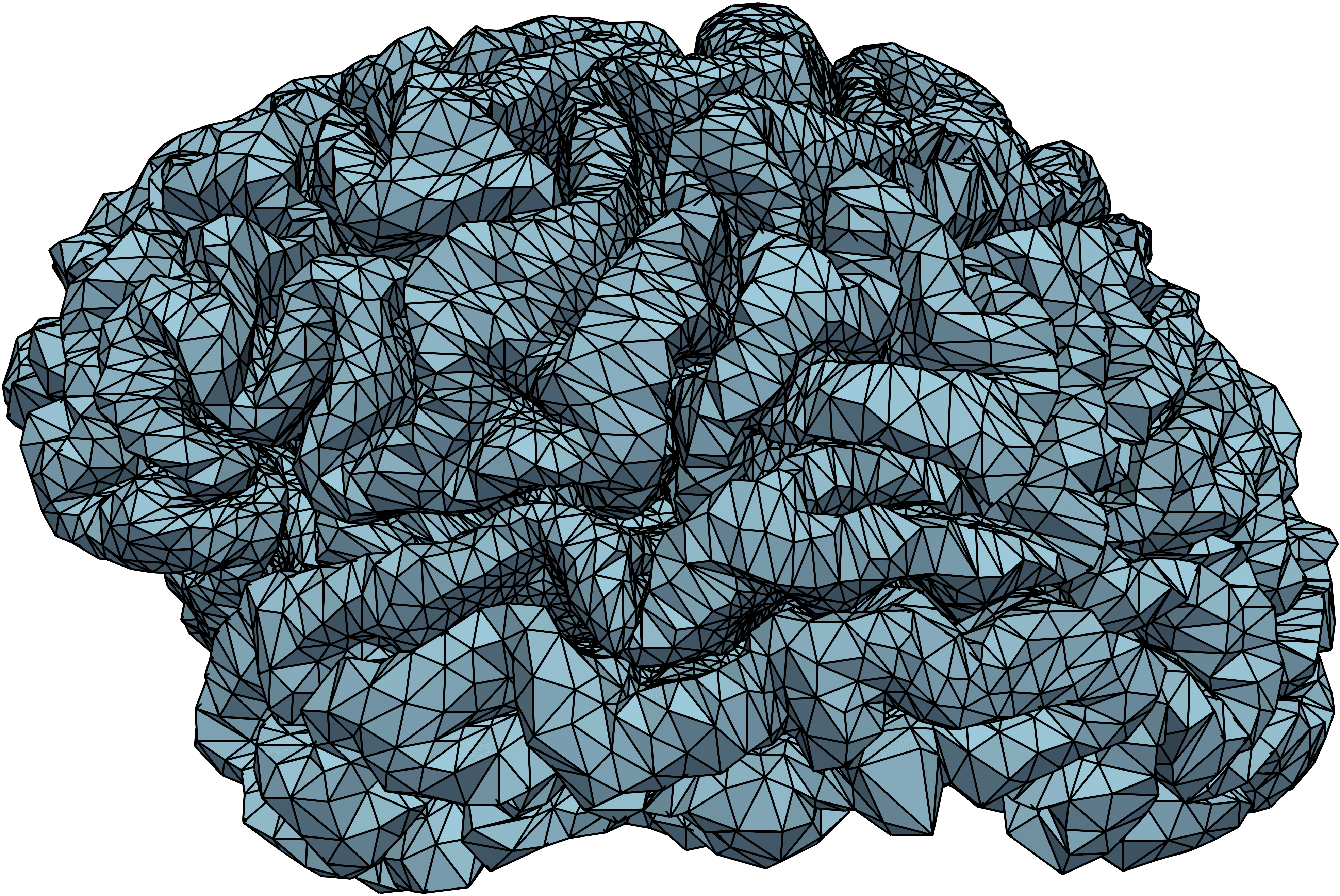}}
        \subfloat[]{
            \includegraphics[width=.24\linewidth]{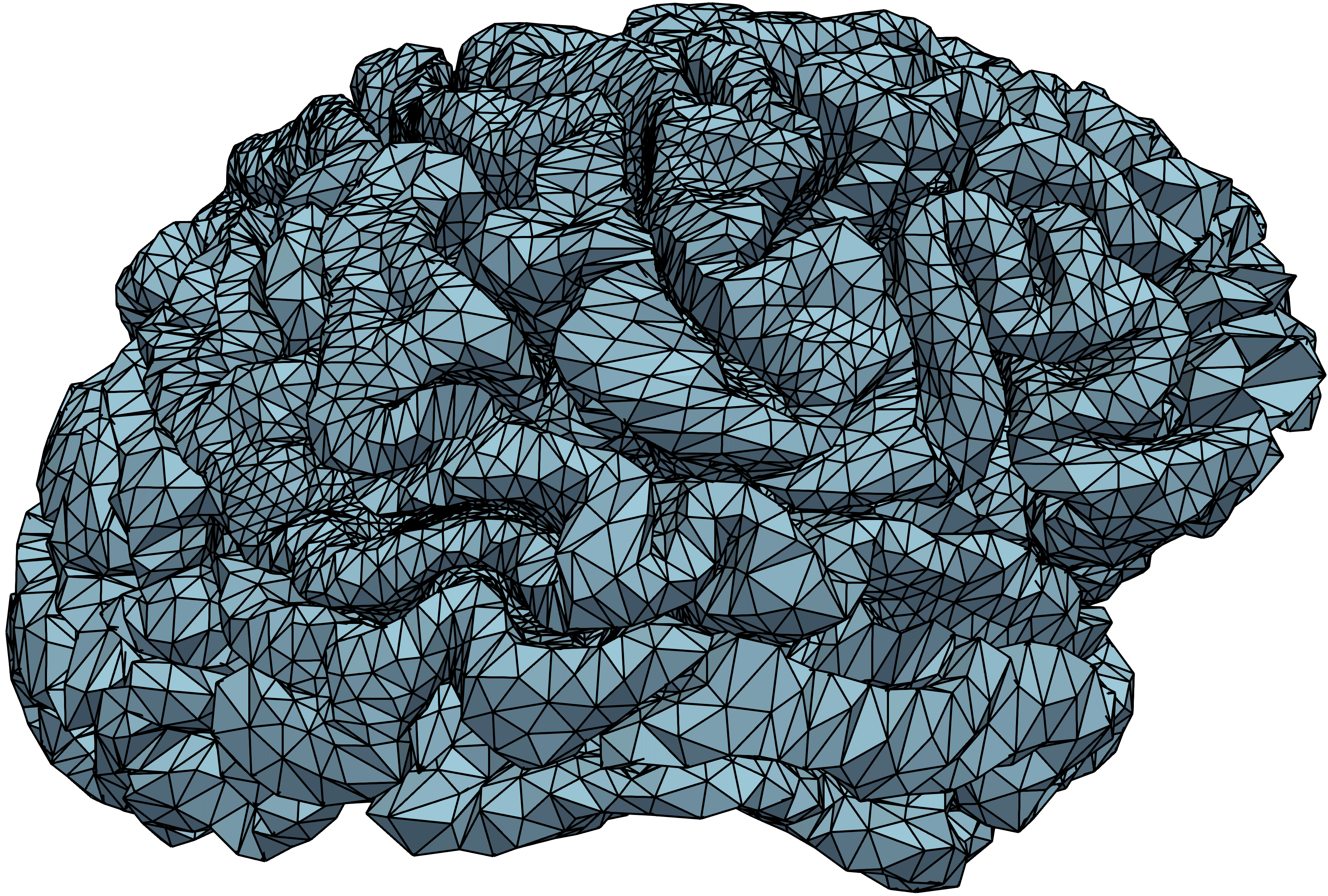}}
        \subfloat[]{
            \includegraphics[width=.24\linewidth]{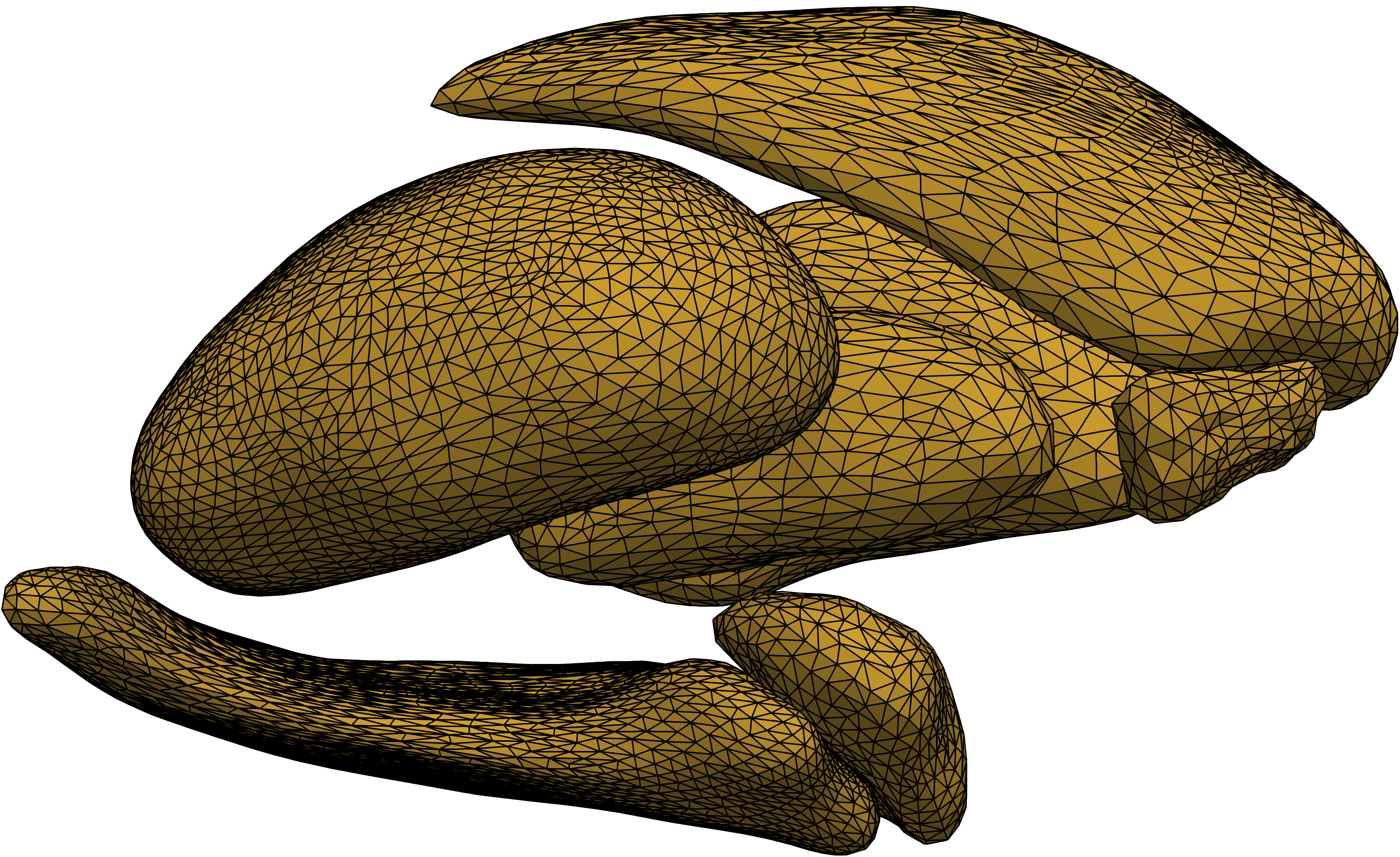}}
        \subfloat[]{
            \includegraphics[width=.24\linewidth]{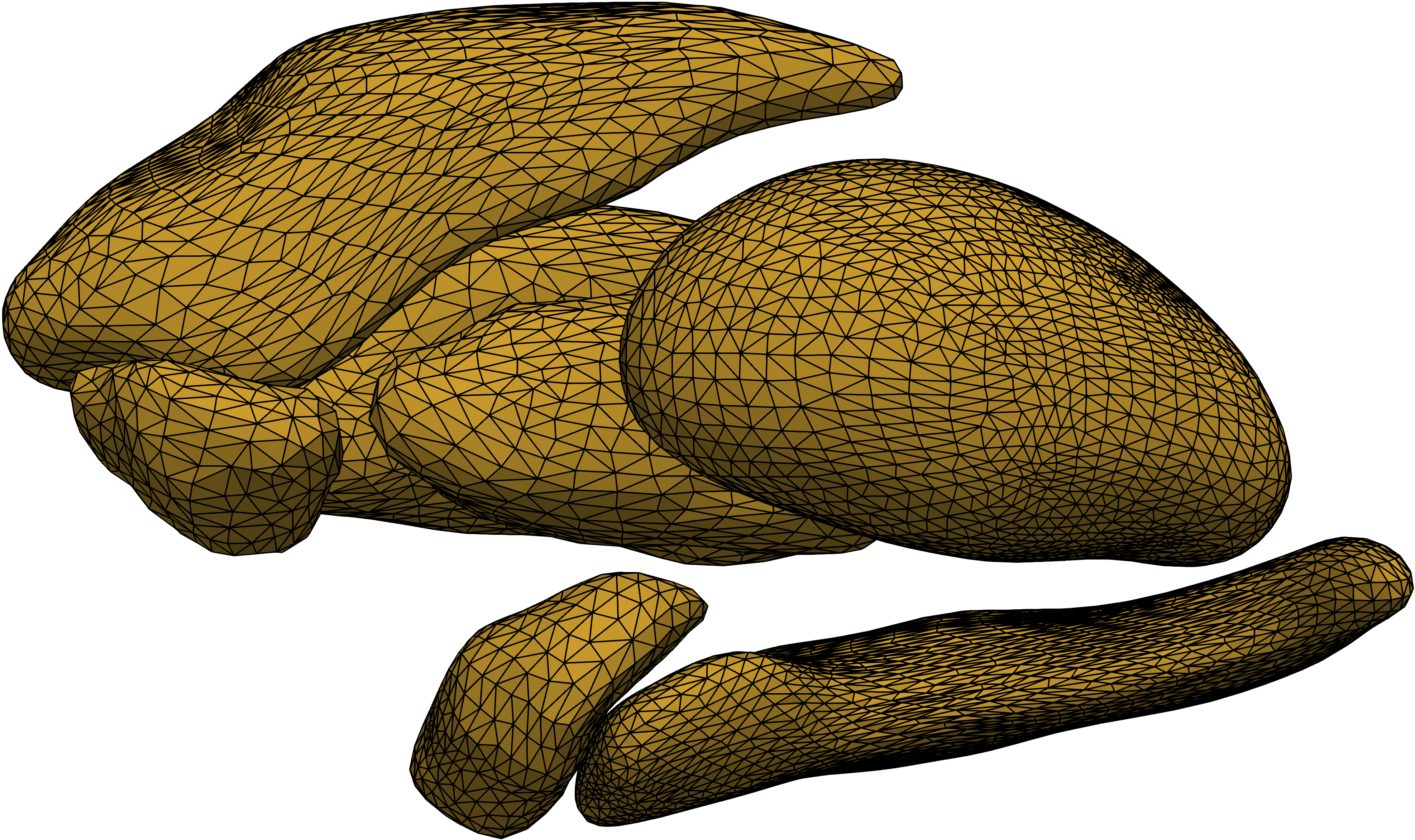}}
        \caption{Cortical meshes from a randomly selected HC subject (blue) and meshes of the subcortical structures from a randomly selected ADD subject (yellow). Presented are lateral views (a-b) of the HC's left hemisphere (LH) and right hemisphere (RH) cortical meshes respectively. Medial views of the ADD subject's LH and RH subcortical structure meshes are also presented (c-d).}
            \label{fig:mni_template}
    \end{figure}

Cortical vertices were assigned 6 features: the $x$, $y$, and $z$ coordinates of both the white matter (WM) and gray matter (GM) vertices in the native space. This was decided because vertices on these surfaces use the same edge weights and therefore the same ``faces'' with different coordinates for the vertices of the respective triangles. Similar to the cortex, subcortical vertices had 3 features: their corresponding $x$, $y$, and $z$ coordinates in the native space as well. To maintain the same number of features for all vertices per scan, the corresponding cortical and subcortical feature matrices were block-diagonalized into a single node feature matrix per scan s.t. $\mathbf{X}\in \mathbb{R}^{47,616 \times 9}$. Sample meshes extracted from a randomly selected HC and one with ADD are demonstrated in Figure \ref{fig:mni_template}.
\subsection{Residual Network Architecture}
\begin{figure}[htb]
        \centering
        \includegraphics[width=.9\linewidth]{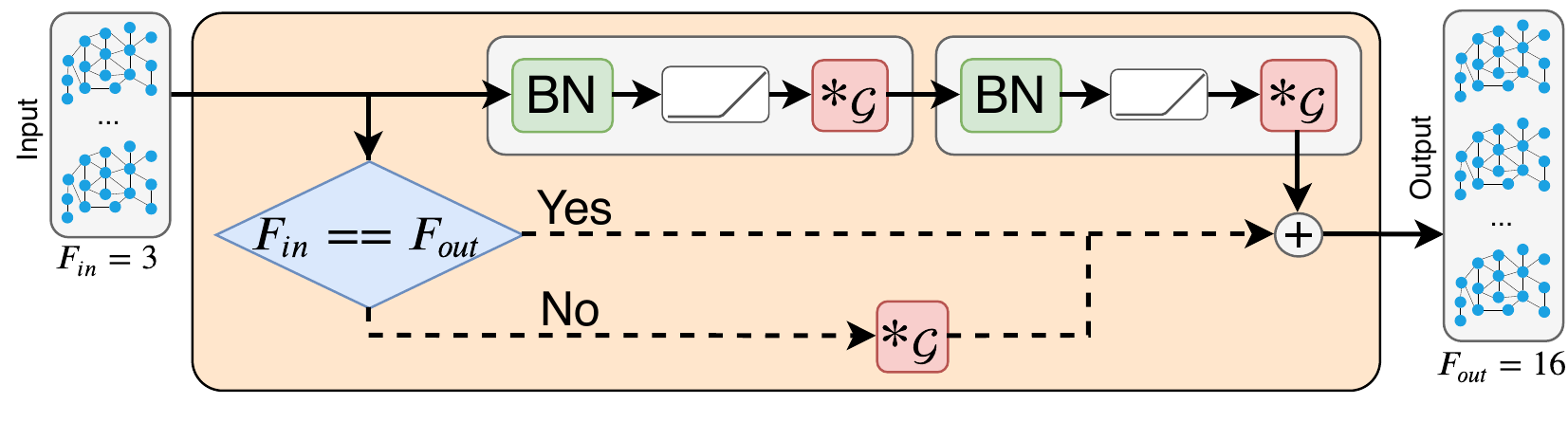}
        \caption{Single ResBlock in the GCN architecture used in this study. Linear mapping of $F_{in}$ to $F_{out}$ channels is implemented using a convolutional layer, $*_\mathcal{G}$. This is done to match the number of input features to the number of desired feature maps.}
        \label{fig:resblock}
    \end{figure}

\begin{figure}[htb]
        \centering
        \includegraphics[width=.9\linewidth]{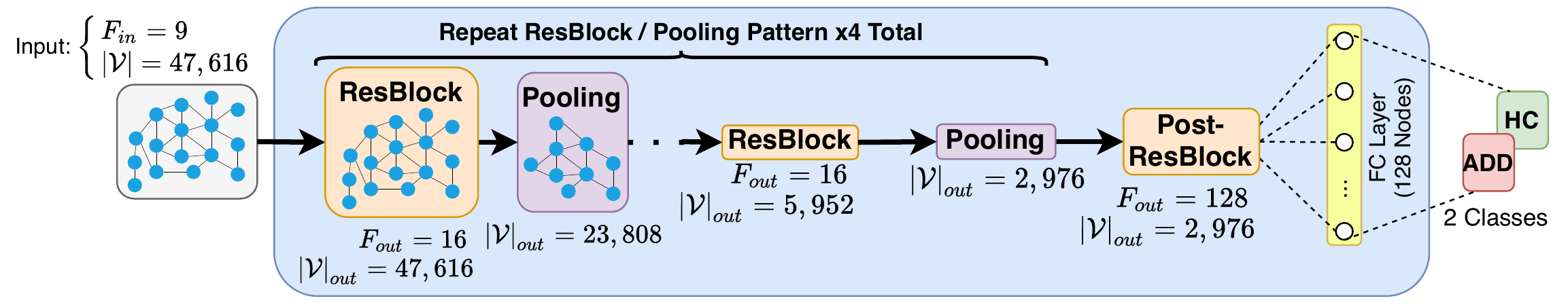}
        \caption{Residual GCN used for the binary classification of ADD. In this study, max-pooling operations are used to downsample the vertex dimension by a factor of 2.}
        \label{fig:total_architecture}
    \end{figure}

Inspired by the work of He \emph{et al.} in \cite{he2016deep}, we propose an improvement upon ChebNet \cite{Defferrard2016} using residual connections within GCNs, which have been shown in prior work to address the common ``vanishing gradient'' problem and improve the performance of deep NNs. Typically, these types of residual networks (ResNets) are implemented by using batch normalization (BN) \cite{pmlr-v37-ioffe15} before a ReLU activation function, and followed by convolution as seen in Fig. \ref{fig:resblock}. Using ResBlocks (Fig. \ref{fig:resblock}), max-pooling operations as described by Defferrard \emph{et al.} \cite{Defferrard2016}, and a standard fully connected (FC) layer \cite{hinton_fully}, the total architecture used in our study is defined in Fig. \ref{fig:total_architecture}. An additional ResBlock, which we refer to as a ``post-ResBlock,'' was introduced prior to the FC layer as a linear mapping tool to match the number of FC units.
\subsection{Grad-CAM Mesh Adaptation}
Interpretability of CNNs was addressed by \cite{Grad-CAM} via their gradient-weighted class activation map (Grad-CAM) approach. In their work, images are fed to CNNs and gradients for each class score (logits prior to softmax) are extracted at the last convolutional layer. Using these gradients, they perform a global average pooling (GAP) operation for each feature map per class to extract ``neuron importance weights,'' $\alpha_c^{(k)} \in \mathbb{R}^{c \times k}$, whose formulation we readapt for meshes s.t.
\begin{equation}
    \alpha_c^{(k)} = \frac{1}{N}\sum_n \frac{\partial y_c}{\partial A_n^{(k)}},
\end{equation}
where $y_c$ corresponds to the class score of class $c$, and $A_n^{(k)}$ refers to the value at vertex $n$ for the $k$-th feature map $A^{(k)} \in \mathbb{R}^N$. A set of neuron importance weights, $\alpha_c^{(k)}$, is extracted for each $k$-th feature map, $A^{(k)}$, and projected onto them to get the class activation maps (CAMs) s.t.
\begin{equation}
    M_c = \textrm{ReLU} \left(\sum_k \alpha_c^{(k)} A^{(k)}\right) \in \mathbb{R}^N.
\end{equation}
As a consequence of pooling, CAMs are upsampled to the same number of nodes as the input mesh for a direct ``overlay'' using a trivial interpolation by going backward along the hierarchical tree used by the pooling operations.
\section{Experimental Design}
\subsection{Dataset \& Preprocessing} \label{sec:dataset}
T1-weighted MRIs from ADNI \cite{Jack2008} were selected with ADD/HC diagnosis labels given up to 2 months after the corresponding scan. This was taken as a precaution to ensure that each diagnosis had clinical justification. The dataset in our study consisted of 1,191 different scans for 435 unique subjects. Section \ref{net_arch_exp} outlines our stratified data splitting strategy to ensure no data leakage occurs at the subject level across the training, validation, and testing sets \cite{Fung2019}.

Meshes for each MRI were extracted following the process described in Section \ref{sec:graph_extract}. The spatial standard deviation from Eq. \ref{eq:edge_weights}, $\sigma$, was set to $2$ ad-hoc. The visual quality for each mesh was assessed manually via a direct overlay over slices of the corresponding MRI. Laplacians for the cortex and each subcortical structure were block-diagonalized to create one overall $\mathbf{L}$ representing a single mesh with multiple connected components. Extracted feature matrices for each sample were min-max normalized per feature to the interval $[-1,1]$ prior to feeding batches of data into the networks. The added zeros during block-diagonalization (as discussed in Section \ref{sec:graph_extract}) were ignored during each normalization step.
\subsection{Network Architecture \& Training} \label{net_arch_exp}
Extra care was taken in the shuffling of samples to avoid bias from subject overlap in our cross-validation \cite{Fung2019}. A custom dataset splitting function was implemented s.t. the distribution of labels was preserved amongst each set while also ensuring to avoid subject overlap. 20\% of the samples were selected at random for the testing set. Of the remaining 80\%, 20\% of those were withheld as the validation set, while the remaining belonged to the training set. A 25-trial Monte Carlo cross-validation was performed using this data split scheme.

The architecture in Fig. \ref{fig:total_architecture} was implemented using 16 kernels per convolutional layer (not including the post-ResBlock), Chebyshev polynomials of order $K=3$, and pooling windows of size $p=2$. Four alternating ResBlock and pooling layers were cascaded as shown in Fig. \ref{fig:total_architecture} prior to the post-ResBlock. The number of units at the post-ResBlock and FC layer was 128. Our GCN was optimized by minimizing a standard binary cross-entropy loss function
    \begin{equation}
        \mathcal{L} = -\frac{1}{N} \sum_{n=1}^N y_n \log(\hat{y_n}) + \left(1-y_n\right)\log\left(1-\hat{y_n}\right),
    \end{equation}
where $\hat{y_n}$ is the predicted class for the $n^{th}$ sample of $N$ total samples and $y_n$ is the ground truth label for the same sample index, $n$.

Networks were trained using batches of 32 samples per step for 100 epochs in each Monte Carlo trial. The Adam \cite{Kingma2015} optimizer was used with a learning rate of $5 \times 10^{-4}$ and a learning rate decay of 0.999. Experiments were implemented in Python 3.6 using Tensorflow 1.13.4 using an NVIDIA GeForce GTX TITAN Z GPU in a Dell Precision Tower 7910 with Linux Mint 19.2.
\section{Results \& Discussion}
\subsection{ADD vs. HC Classification}
\begin{figure}[!htb]
        \centering
        \includegraphics[width=0.75\linewidth]{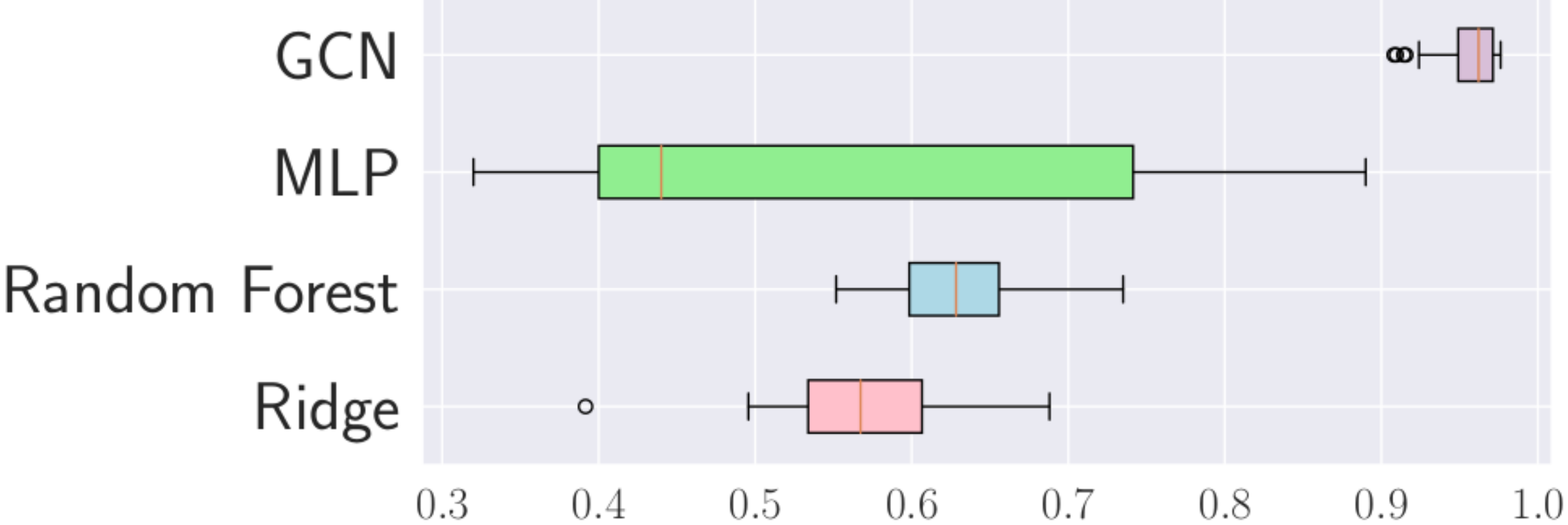}
        \caption{Monte Carlo cross-validation accuracy results for GCN and baseline model architectures from \cite{PARISOT2018117} used on brain meshes.}
        \label{fig:box_plots}
    \end{figure}

Our cross-validation includes the same multilayer perceptron (MLP) classifier architecture, ridge classifier, and a 100-estimator random forest classifier set up by Parisot \emph{et al.} in \cite{PARISOT2018117}, where a similar graph approach is also used on the classification of ADD based on population graphs. The MLP designed was synonymous to the design in \cite{PARISOT2018117} s.t. the number of hidden layers and parameters was the same as our GCNs. Demonstrated in Figure \ref{fig:box_plots}, our GCN outperformed other standard classifiers not limited to graph methods on our dataset split.
    \begin{table}[!htb]
        \caption{Model comparison to classifiers in studies not limited to surface methods.}
        \begin{tabular}{ccccccc}
            \hline
            \textbf{Study} & \textbf{Data} & \textbf{ADD/HC} & \textbf{Acc. (\%)}
            & \textbf{Sens. (\%)}
            & \textbf{Spec. (\%)} & \textbf{AUC (\%)}\\
            \hline
            \cite{arjun2019plosone} & MRI & --/-- (723) & 73.76 & -- & -- & --\\
            \cite{arjun2019plosone} & MRI+PET$_\text{amyloid}$ & --/-- (723)  & 92.34 & -- & -- & --\\
            \cite{LIU2014466} & MRI+PET$_\text{FDG}$ & 51/52 & 94.37 & 94.71 & 94.04 & \textbf{97.24}\\
            \cite{WESTMAN2012229} & MRI+CSF & 96/111 & 91.80 & 88.50 & 94.60 & 95.80\\
            \cite{HU2016132} & MRI & 228/188 & 84.13 & 82.45 & 85.63 & 90.00\\
            \cite{beheshti2017classification} & MRI & 92/94 & 93.01 & 89.13 & \textbf{96.80} & 93.51\\
            \cite{hosseini2016alzheimer} & MRI & 70/70 & \textbf{97.60} & -- & -- & --\\
            \cite{gupta2013natural} & MRI & 200/232 & 94.74 & \textbf{95.24} & 94.26 & --\\
            Ours & MRI & 167/265 & 96.35 & 92.37 & 96.74 & 96.84\\
            \hline
        \end{tabular}
        \label{table:compare_other}
    \end{table}
    
The results in Table \ref{table:compare_other} highlight comparable metrics of our model versus other studies that operate on voxels from full 3D MRI volumes, including \cite{arjun2019plosone}. In their work, Punjabi \emph{et al.} train a multi-modal CNN using both volumetric MRI and FDG-PET imaging for the same task, which we outperform while training and evaluating on a smaller subset of their subject population. Furthermore, volumetric models like those in \cite{arjun2019plosone} rely on 3D CNNs with far more learned parameters, e.g. \cite{arjun2019plosone}'s 200,194,502 weights ($\times 2$ for fusion model), in comparison to our GCN's 497,522 learned parameters needed for comparable results. Like \cite{ranjan2018generating}, we also achieve comparable results with far less learning parameters by working on meshes and focusing on brain \emph{shape} instead of working on raw voxels obtained from MRIs and using voxel-based approaches.
\subsection{Class Activation Map Visualization}
\begin{figure}[htb]
        \centering
        \subfloat[]{
            \includegraphics[width=.24\linewidth]{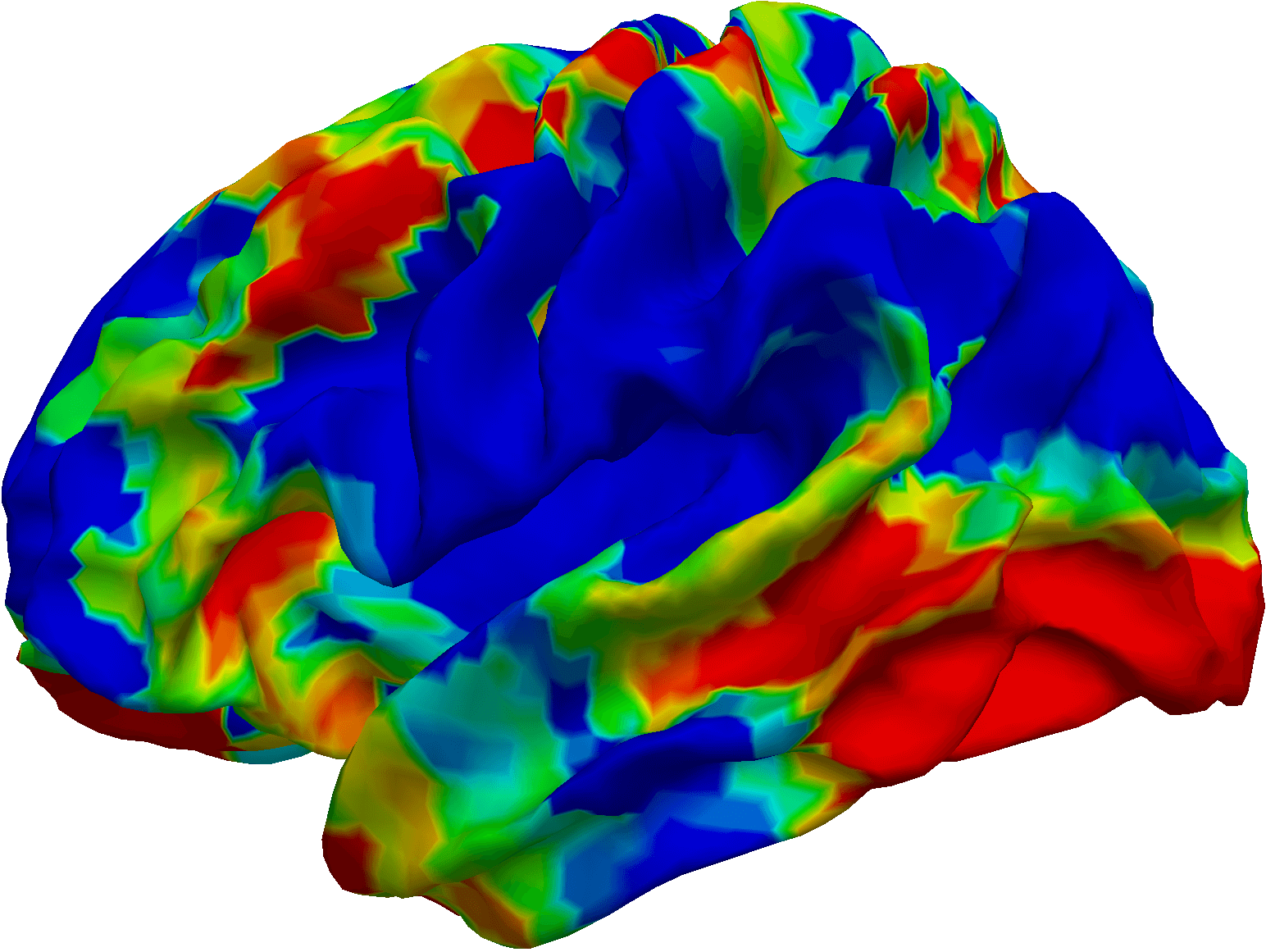}}
        \subfloat[]{
            \includegraphics[width=.24\linewidth]{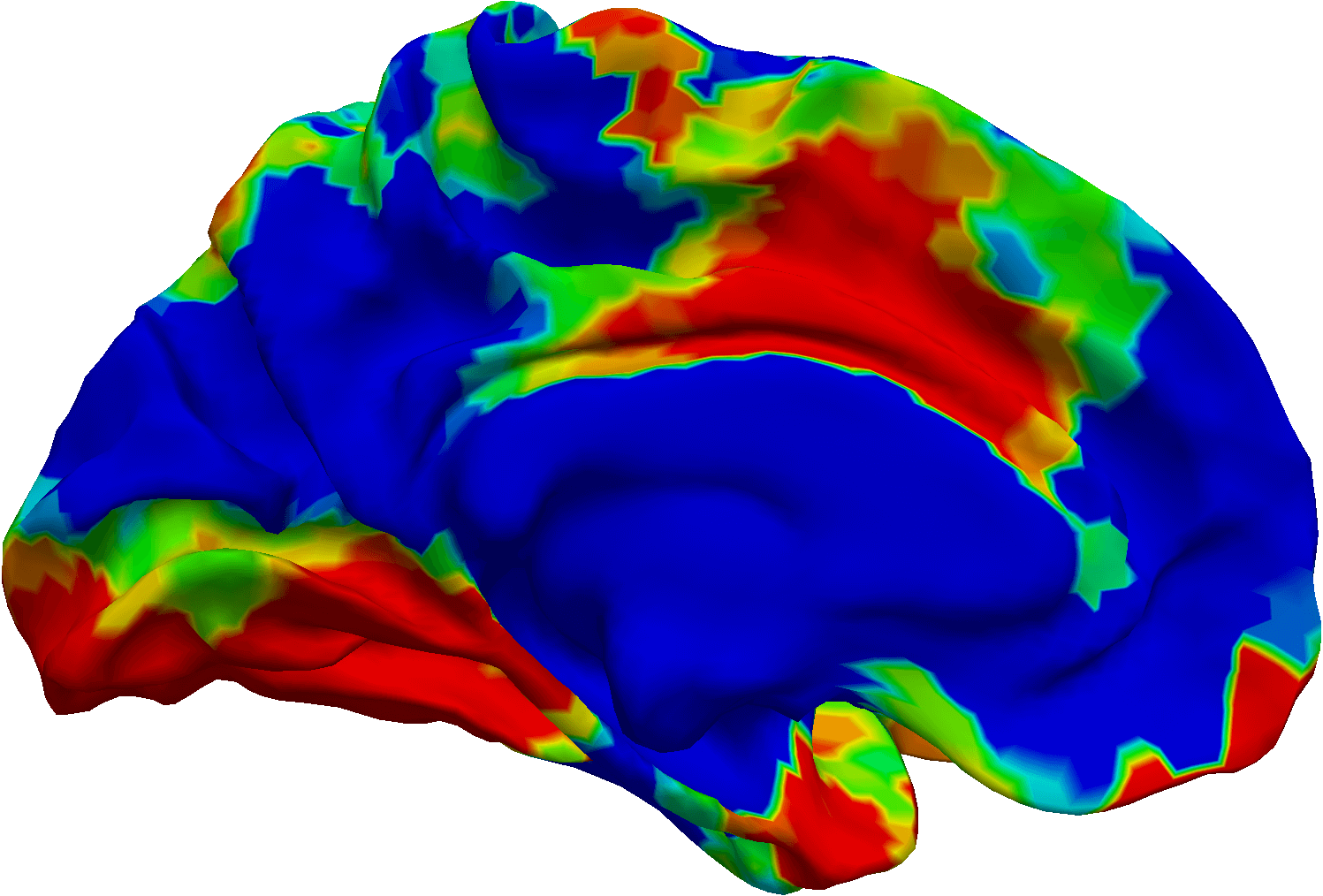}}
        \subfloat[]{
            \includegraphics[width=.24\linewidth]{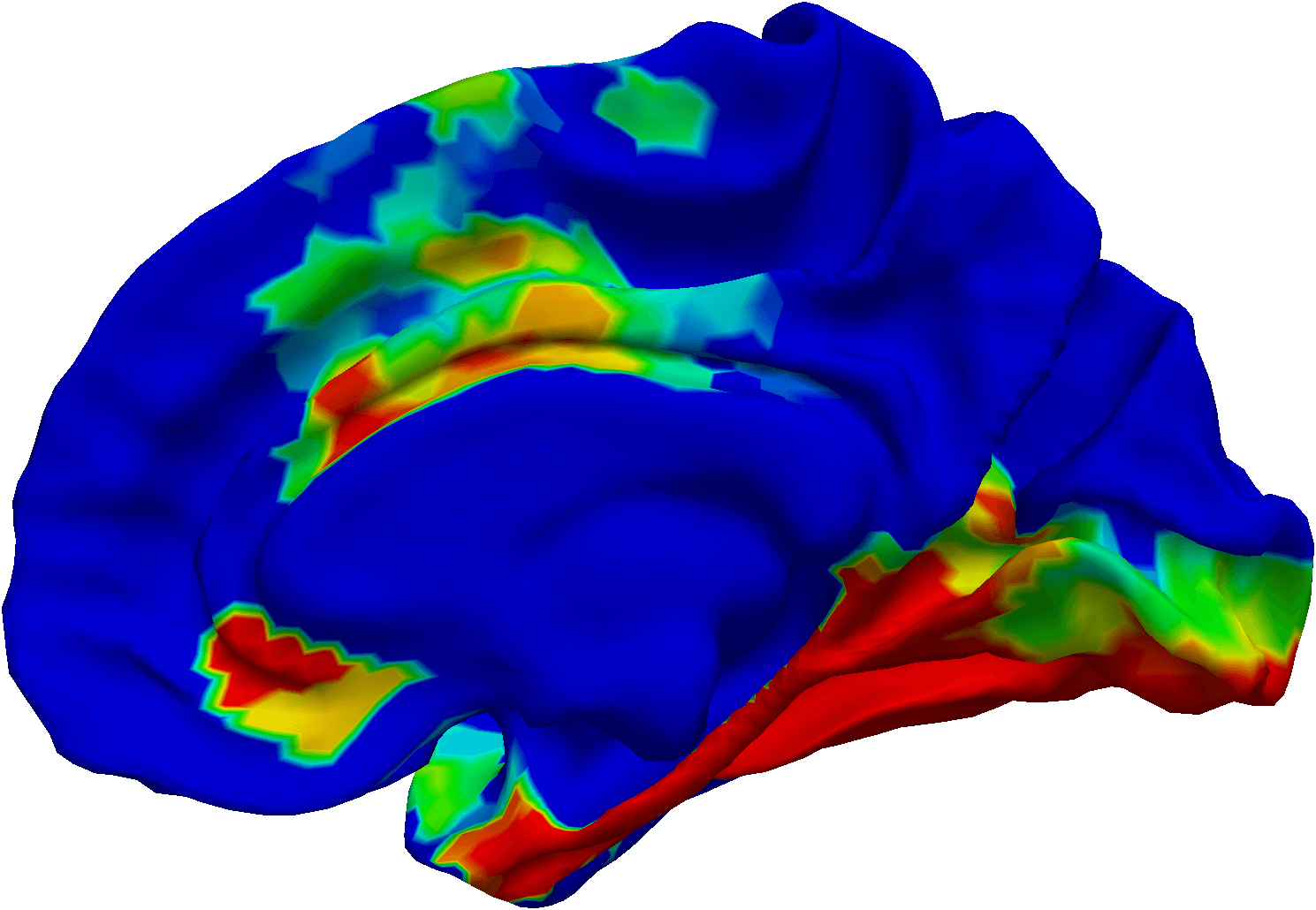}}
        \subfloat[]{
            \includegraphics[width=.24\linewidth]{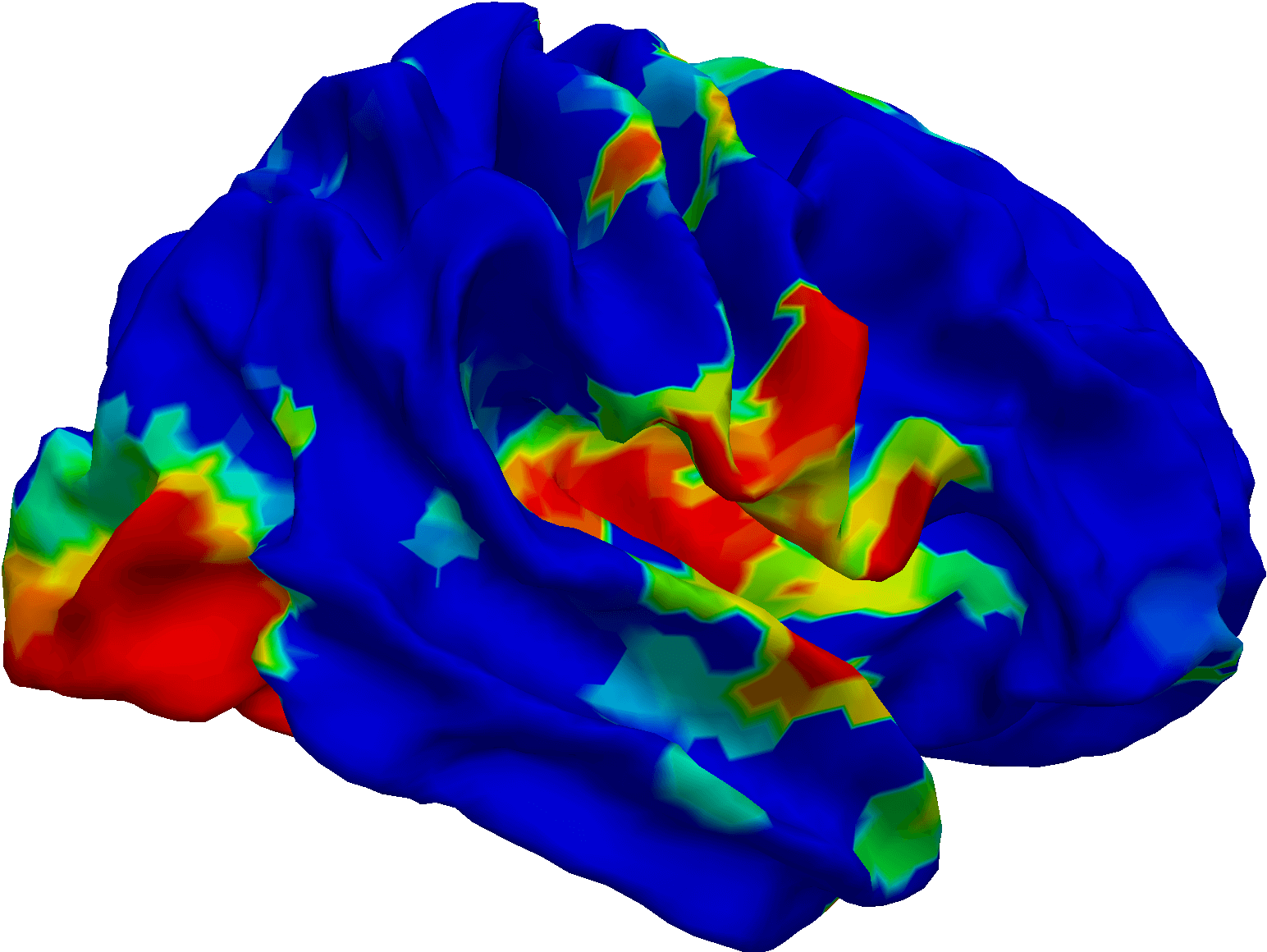}}
        \newline
        \subfloat[]{
            \includegraphics[width=.24\linewidth]{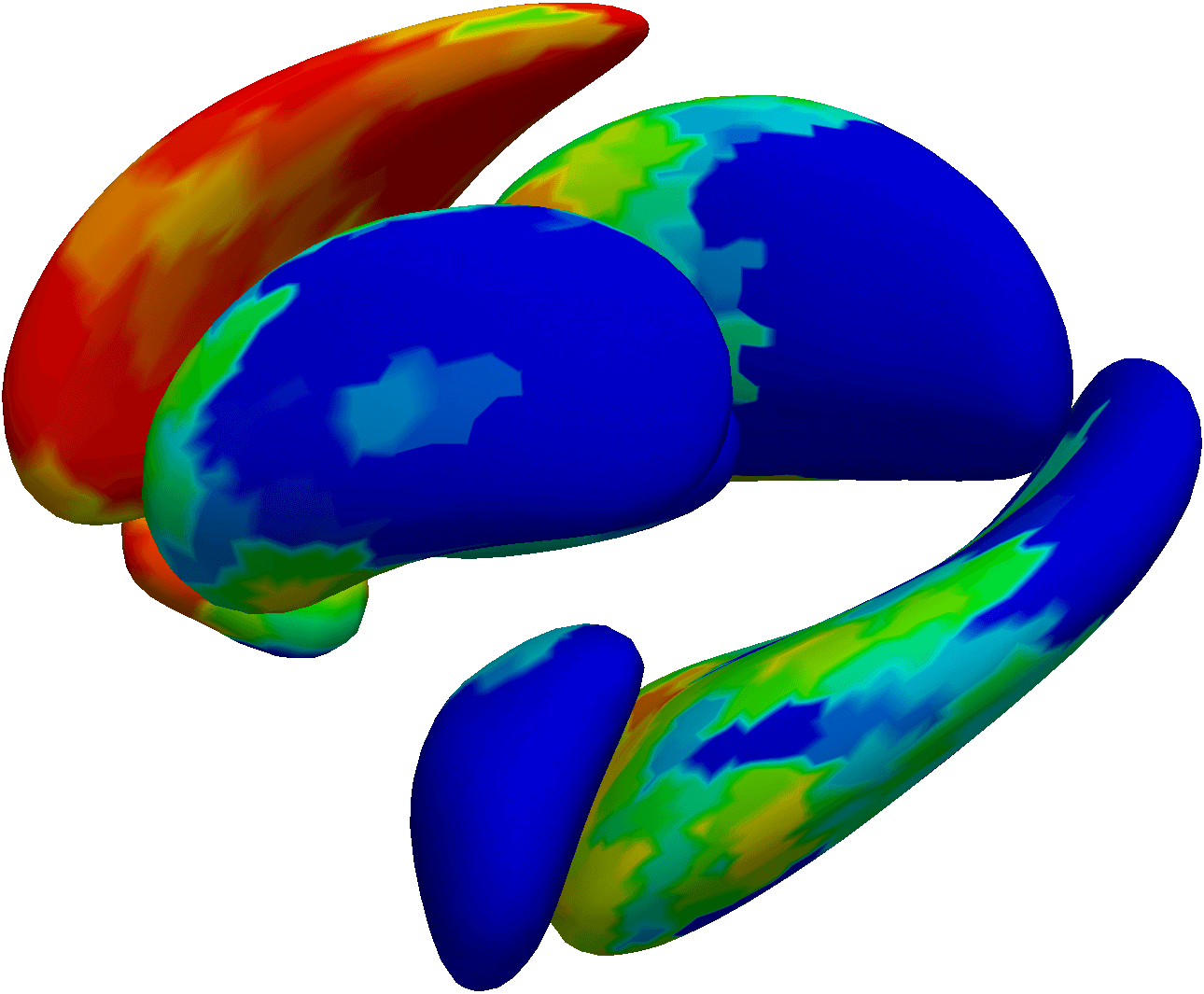}}
        \subfloat[]{
            \includegraphics[width=.24\linewidth]{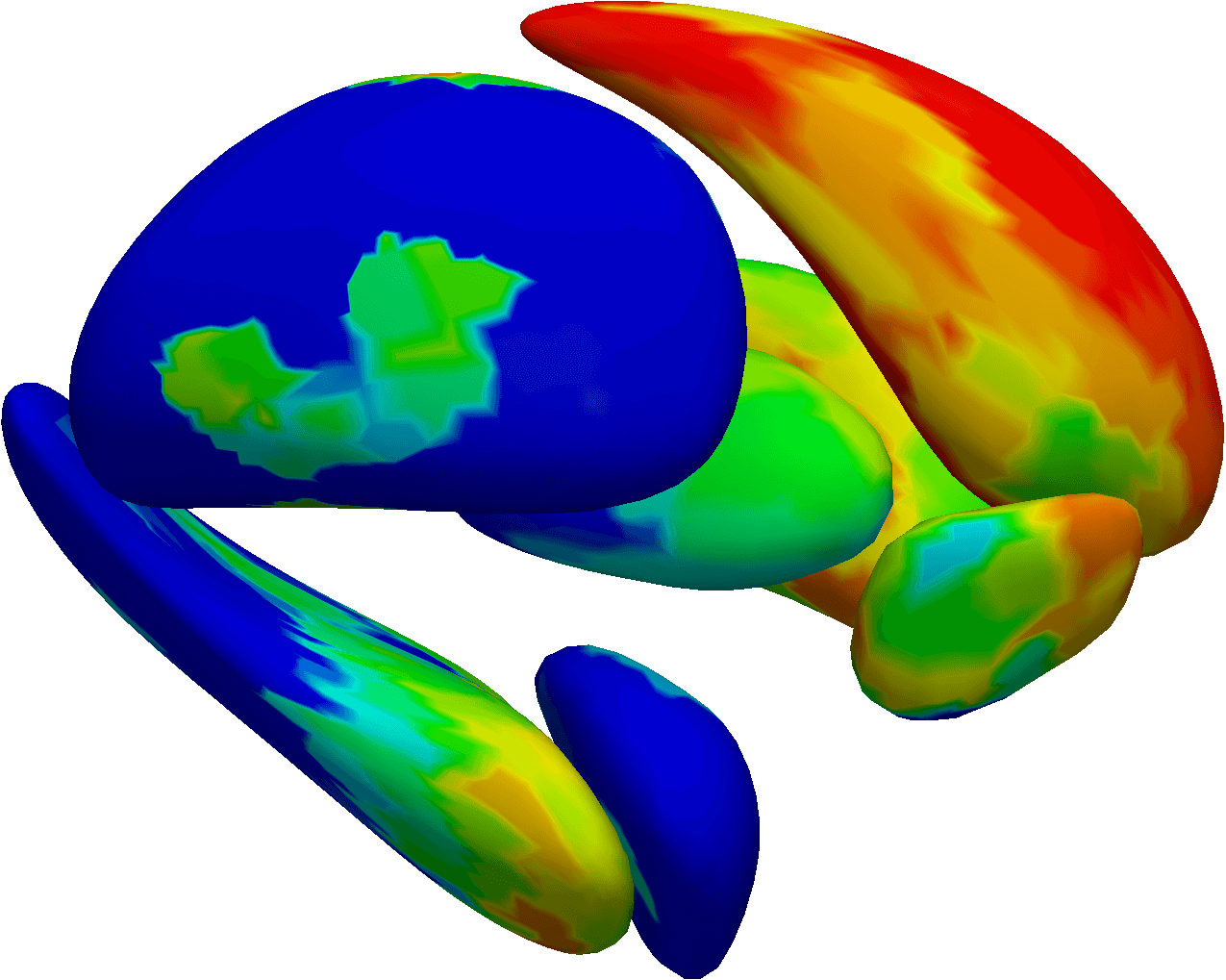}}
        \subfloat[]{
            \includegraphics[width=.24\linewidth]{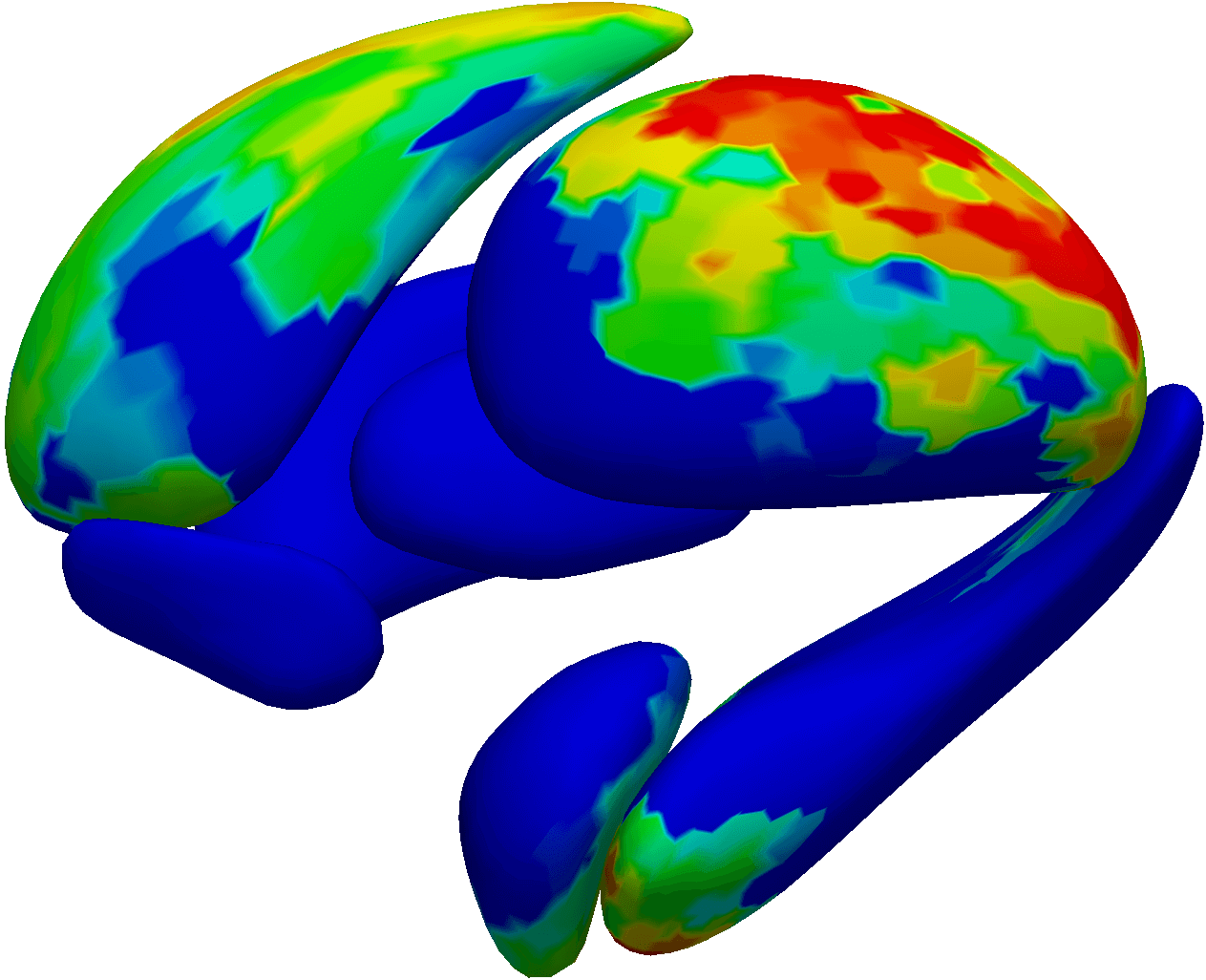}}
        \subfloat[]{
            \includegraphics[width=.24\linewidth]{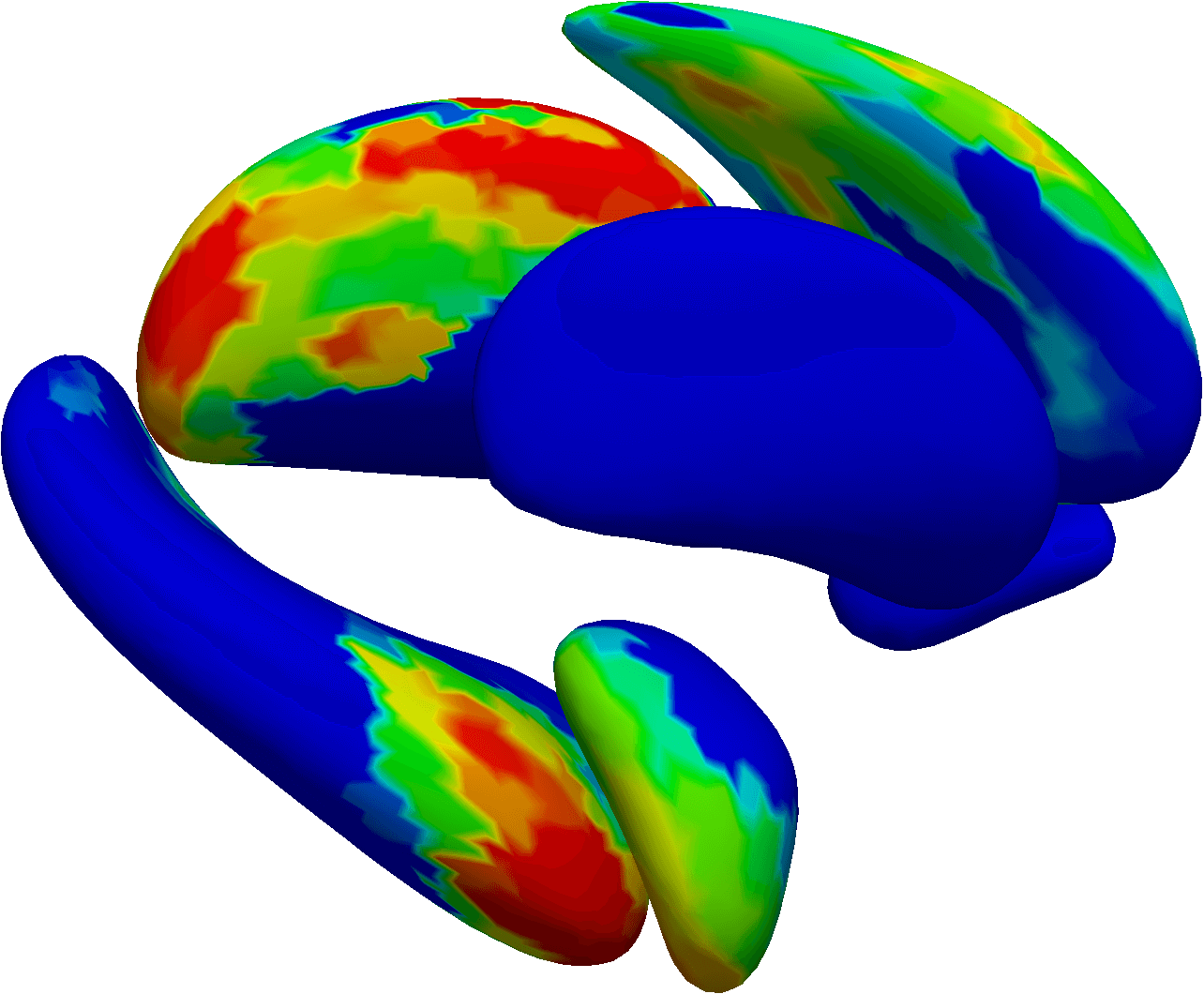}}
        \newline
        \subfloat{
            \includegraphics[width=.24\linewidth]{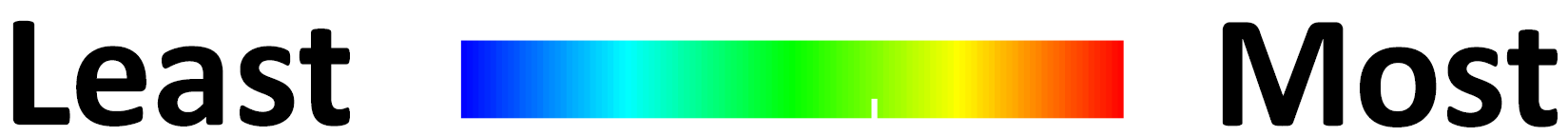}}
        \caption{Average TP CAMs on the cortical template from \cite{fischl1999high, Fischl2012} (top) and subcortical structures from \cite{BESSON2014283} (bottom) including: (a-b, e-f) lateral-medial views of the LH respectively, (c-d, g-h) medial-lateral views of the RH respectively.}
        \label{fig:grad_cam_AD}
        \centering
    \end{figure}

By employing Grad-CAM on our best GCN, an average CAM was generated for true positive (TP) predictions (Fig. \ref{fig:grad_cam_AD}). We project our CAM onto the cortical template \cite{fischl1999high} provided by FreeSurfer \cite{Fischl2012} and the homemade subcortical structure templates detailed in \cite{BESSON2014283}. The color scale highlights areas from least-to-most influential in TP predictions. The patterns in the CAM match previously described distributions of cortical and subcortical atrophy \cite{Kalin2017a, Dickerson2009}. One reason we may observe a mismatch between the CAM and expected atrophy in the inferior parietal lobule could be the degree of variability in highly folded association cortex, e.g., the intermediate sulcus of Jensen is found only in some individuals \cite{ono1990atlas, Thompson1998}. The slightly more left lateralized pattern in the CAM aligns with previous reports that propose greater pathologic burden and neurodegeneration of the language network which leads to worsening on verbal-based neuropsychological measures of memory resulting in a diagnosis for ADD \cite{Derflinger2011}.
    \newcommand{\imsize}{0.23\textwidth}
\section{Conclusion \& Future Work}
In this work, we demonstrated the effectiveness of using cortical and subcortical surface meshes in the context of binary ADD clinical diagnosis and ROI visualization in TP predictions. Furthermore, we compared the cross-validation results of our model for the same ADD vs. HC problem using other ML models on our data. Additionally, our final results were comparable to the results of other studies that use traditional neuroimaging modalities as inputs. When compared to the performance of the multimodal approach used in \cite{arjun2019plosone}, our model outperforms their approach, thus potentially indicating the reliability of leveraging shape information represented as meshes to perform the same binary classification task.

Natural extensions of this work could be to (1) expand our classification problem to include a third class from ADNI, mild cognitive impairment (MCI), (2) increase the population in our study to include those in ADNI3 \cite{Jack2008}, (3) work on longitudinal predictions, and (4) compare our model's performance in using only the cortex, subcortical structures, or both. Additionally, having a 3D-volume-to-mesh dataset offers the potential for developing generative networks, as in \cite{Goodfellow2014}, for performing the graph extraction preprocessing step described in Section \ref{sec:graph_extract}. This will provide more autonomy and limit the need for the manual quality assessment (QA) of meshes as a part of our pipeline.
\section{Acknowledgements}
This work was funded in part by the Biomedical Data Driven Discovery Training Grant from the National Library of Medicine (5T32LM012203) through Northwestern University, and the National Institute on Aging. The authors would also like to thank the QUEST High Performance Computing Cluster at Northwestern University for computational resources.

Data collection and sharing for this project was funded by the Alzheimer’s Disease Neuroimaging Initiative (ADNI). Data collection and sharing for this project was funded by the Alzheimer's Disease Neuroimaging Initiative (ADNI) (National Institutes of Health Grant U01 AG024904) and DOD ADNI (Department of Defense award number W81XWH-12-2-0012). ADNI is funded by the National Institute on Aging, the National Institute of Biomedical Imaging and Bioengineering, and through generous contributions from the following: AbbVie, Alzheimer’s Association; Alzheimer’s Drug Discovery Foundation; Araclon Biotech; BioClinica, Inc.; Biogen; Bristol-Myers Squibb Company; CereSpir, Inc.; Cogstate; Eisai Inc.; Elan Pharmaceuticals, Inc.; Eli Lilly and Company; EuroImmun; F. Hoffmann-La Roche Ltd and its affiliated company Genentech, Inc.; Fujirebio; GE Healthcare; IXICO Ltd.; Janssen Alzheimer Immunotherapy Research \& Development, LLC.; Johnson \& Johnson Pharmaceutical Research \& Development LLC.; Lumosity; Lundbeck; Merck \& Co., Inc.; Meso Scale Diagnostics, LLC.; NeuroRx Research; Neurotrack Technologies; Novartis Pharmaceuticals Corporation; Pfizer Inc.; Piramal Imaging; Servier; Takeda Pharmaceutical Company; and Transition Therapeutics. The Canadian Institutes of Health Research is providing funds to support ADNI clinical sites in Canada. Private sector contributions are facilitated by the Foundation for the National Institutes of Health (\url{https://www.fnih.org}). The grantee organization is the Northern California Institute for Research and Education, and the study is coordinated by the Alzheimer’s Therapeutic Research Institute at the University of Southern California. ADNI data are disseminated by the Laboratory for Neuro Imaging at the University of Southern California.
\bibliographystyle{splncs04}
\bibliography{mybibliography}
\end{document}